\documentstyle[epsfig]{mn}
\newif\ifAMStwofonts
\newcommand{\be}{\begin{equation}}    
\newcommand{\ee}{\end{equation}}
\newcommand{\beq}{\begin{eqnarray}}
\newcommand{\eeq}{\end{eqnarray}}

\def\nn{\nonumber}
\def\op{ $ }
\def\cl{$ }

\def\msun{M_\odot}

\ifoldfss
  \ifCUPmtlplainloaded \else
    \NewTextAlphabet{textbfit} {cmbxti10} {}
    \NewTextAlphabet{textbfss} {cmssbx10} {}
    \NewMathAlphabet{mathbfit} {cmbxti10} {} 
    \NewMathAlphabet{mathbfss} {cmssbx10} {} 
  \fi
  \ifAMStwofonts
    \ifCUPmtlplainloaded \else
      \NewSymbolFont{upmath} {eurm10}
      \NewSymbolFont{AMSa} {msam10}
      \NewMathSymbol{\upi}     {0}{upmath}{19}
      \NewMathSymbol{\umu}     {0}{upmath}{16}
      \NewMathSymbol{\upartial}{0}{upmath}{40}
      \NewMathSymbol{\leqslant}{3}{AMSa}{36}
      \NewMathSymbol{\geqslant}{3}{AMSa}{3E}

      \let\leq=\leqslant 
      \let\geq=\geqslant 
    \fi
  \fi
\fi 

\ifnfssone
  \newmathalphabet{\mathit}
  \addtoversion{normal}{\mathit}{cmr}{m}{it}
  \addtoversion{bold}{\mathit}{cmr}{bx}{it}
  \newmathalphabet{\mathbfit} 
  \addtoversion{normal}{\mathbfit}{cmr}{bx}{it}
  \addtoversion{bold}{\mathbfit}{cmr}{bx}{it}
  \newmathalphabet{\mathbfss} 
  \addtoversion{normal}{\mathbfss}{cmss}{bx}{n}
  \addtoversion{bold}{\mathbfss}{cmss}{bx}{n}
  \ifAMStwofonts
    \ifCUPmtlplainloaded \else
      \UseAMStwoboldmath
      \makeatletter
      \new@mathgroup\upmath@group
      \define@mathgroup\mv@normal\upmath@group{eur}{m}{n}
      \define@mathgroup\mv@bold\upmath@group{eur}{b}{n}
      \edef\UPM{\hexnumber\upmath@group}
      \new@mathgroup\amsa@group
      \define@mathgroup\mv@normal\amsa@group{msa}{m}{n}
      \define@mathgroup\mv@bold\amsa@group{msa}{m}{n}
      \edef\AMSa{\hexnumber\amsa@group}
      \makeatother
      \mathchardef\upi="0\UPM19
      \mathchardef\umu="0\UPM16
      \mathchardef\upartial="0\UPM40
      \mathchardef\leqslant="3\AMSa36
      \mathchardef\geqslant="3\AMSa3E

      \let\leq=\leqslant 
      \let\geq=\geqslant 
    \fi
  \fi
\fi 

\ifnfsstwo
  \DeclareMathAlphabet{\mathbfit}{OT1}{cmr}{bx}{it}
  \SetMathAlphabet\mathbfit{bold}{OT1}{cmr}{bx}{it}
  \DeclareMathAlphabet{\mathbfss}{OT1}{cmss}{bx}{n}
  \SetMathAlphabet\mathbfss{bold}{OT1}{cmss}{bx}{n}
  \ifAMStwofonts
    \ifCUPmtlplainloaded \else
      \DeclareSymbolFont{UPM}{U}{eur}{m}{n}
      \SetSymbolFont{UPM}{bold}{U}{eur}{b}{n}
      \DeclareSymbolFont{AMSa}{U}{msa}{m}{n}
      \DeclareMathSymbol{\upi}{0}{UPM}{"19}
      \DeclareMathSymbol{\umu}{0}{UPM}{"16}
      \DeclareMathSymbol{\upartial}{0}{UPM}{"40}
      \DeclareMathSymbol{\leqslant}{3}{AMSa}{"36}
      \DeclareMathSymbol{\geqslant}{3}{AMSa}{"3E}

      \let\leq=\leqslant 
      \let\geq=\geqslant 
    \fi
  \fi
\fi 

\ifCUPmtlplainloaded \else
  \ifAMStwofonts \else 
    \def\upi{\pi}
    \def\umu{\mu}
    \def\upartial{\partial}
  \fi
\fi

\title{Gravitational Wave Background from a Cosmological 
Population of Core-Collapse Supernovae} 
\author[Valeria Ferrari, Sabino Matarrese and Raffaella Schneider]
       {Valeria Ferrari$^1$, Sabino Matarrese$^2$ and Raffaella Schneider$^3$\\
        $^1$Dipartimento di Fisica ``G. Marconi", 
Universit\'a degli Studi di Roma, ``La Sapienza" 
and Sezione INFN  ROMA1,\\ p.le A.  Moro
5, 00185 Roma, Italy\\
        $^2$Dipartimento di Fisica ``Galileo Galilei",
 Universit\'a degli Studi di Padova
and Sezione INFN  PADOVA,\\ via Marzolo 8, 35131 Padova, Italy\\
        $^3$Dipartimento di Fisica ``E. Fermi", Universit\'a degli Studi di 
Roma, ``La Sapienza"
and Sezione INFN  ROMA1,\\ p.le A.Moro 5, 00185 Roma, Italy} 
\date{April  1998}

\pagerange{\pageref{firstpage}--\pageref{lastpage}}
\pubyear{1998}

\begin{document}

\maketitle

\label{firstpage}

\begin{abstract}
We analyse the stochastic background of gravitational radiation emitted
by a cosmological population of core-collapse supernovae.
The supernova rate as a function of redshift is deduced from 
an observation-based determination of the star formation rate density 
evolution.
We then restrict our analysis to the range of progenitor masses leading to
black hole collapse. In this case, the main features of the
gravitational-wave emission spectra have been shown to be, to some extent, 
independent of the initial conditions and of the equation of state
of the collapsing star, and to depend only on the black hole mass and 
angular momentum.
We calculate the overall signal produced by the ensemble of black-hole
collapses throughout the Universe, assuming a flat cosmology with vanishing 
cosmological constant. 
Within a wide range of parameter values, we find that the
spectral strain amplitude has a maximum at a few hundred Hz with
an amplitude between $10^{-28}$ and $10^{-27}$ Hz$^{-1/2}$; 
the corresponding closure density, $\Omega_{GW}$, has a maximum amplitude 
ranging between $10^{-11}$ and $10^{-10}$ 
in the frequency interval $\sim 1.5-2.5$ kHz.
Contrary to previous claims, our observation-based determination leads to 
a duty cycle of order $0.01$, making our stochastic backgound a non-continuous
one. Although the amplitude of our background is comparable to the 
sensitivity that can be reached by a pair of advanced LIGO detectors, the 
characteristic shot-noise
structure of the predicted signal might be in principle exploited to design 
specific detection strategies.

\end{abstract}
\begin{keywords}
gravitational wave background -- star formation rate: black holes.
\end{keywords}

\section{Introduction}

According to the theory of general relativity, gravitational 
waves are expected to be produced in such a large variety
of astrophysical and cosmological phenomena that it
is plausible to guess that our universe is pervaded by  a
background of gravitational radiation.
Depending on their origin, different contributions to this
background will exhibit peculiar spectral properties,
the features of which it is interesting to investigate 
in view of a possible detection by future 
gravitational antennas.

Gravitational waves of cosmological origin could be the 
result of processes that developed in the very early Universe. 
Due to the extremely small graviton cross-section,
the spectral properties 
of such a relic radiation should have been retained
with no substantial alteration until today, and, if observed, 
would provide information on the physical conditions of the epochs 
when these waves were produced (see, for a recent review, Maggiore 1998). 

Further contributions to the stochastic background 
of gravitational waves are  of astrophysical nature
and their generation  dates back to
more recent epochs, when galaxies and stars started to form and evolve.
For instance, the background generated by the 
radiation emitted  from rotating neutron
stars in the Galaxy has been considered in (Giampieri 1997; Postnov 1997) 
and it has been shown that, by taking into account the 
uncertainty with which the ellipticity of neutron stars 
is known, $h^2\Omega_{GW}(\nu=100~\mbox{Hz})$ is in the range $10^{-15} - 
10^{-9}$,
where \op \Omega_{GW}\cl is the closure energy density of gravitational 
waves per logarithmic frequency interval, 
\be
\Omega_{GW}=\frac{1}{\rho_c}\frac{d\rho_{GW}}{d\log{\nu}},
\ee 
with \op \rho_{c}= 3H_0^2\,/\,8\pi G\cl
the critical density.\footnote{
Throughout this paper we shall write the Hubble constant
as\[H_0=h \times 100\, \mbox{km}~\mbox{s}^{-1}\mbox{Mpc}^{-1}.\]}
A strategy for detecting these signals with one interferometric antenna,
by squaring the detector amplitude and searching for a sidereal modulation
has been proposed in (Giazotto, Bonazzola \& Gourgoulhon 1997; Giampieri 1997).
Further studies have estimated the intensity of 
the stochastic background
produced by the galactic merging of unresolved binary white dwarfs 
(Postnov 1997), and subsequently extended to the extragalactic white dwarf's 
merging for different  cosmological models (Kosenko \& Postnov 1998).
The considered range  of frequency is 
\op 10^{-3} - 10^{-2} \, \mbox{Hz}.\cl

In this paper,  we study the 
background of gravitational waves emitted by sufficiently
massive stars that, having reached the final stages of their evolution, 
collapse to form a black hole. 
We consider only collapses to a black hole 
because numerical studies have shown that 
the spectrum of the gravitational energy emitted in these processes
exhibits some distinctive features that are to some extent
independent of the initial conditions and of the equation of state
of the collapsing star, and that 
depend only on the black hole mass and on its angular
momentum (see, for a recent review, Ferrari \& Palomba 1998).
Conversely,  the simulations
of gravitational collapse to a neutron star
available  in the literature
indicate that the energy spectrum which is produced
strongly depends on the equation of state of the collapsing star,
and no distinctive features can easily be extracted
to model the process.

The  energy spectrum which we will use to model the 
collapse to a black hole is that obtained by Stark \& 
Piran (1985, 1986), who studied the evolution
of the axisymmetric collapse of polytropic, rotating stars,
by a  fully relativistic numerical simulation.

In order to evaluate the contribution that
each burst of gravitational radiation emitted in a collapse
occurred in recent or past epochs
gives to the stochastic background,
the knowledge of the  star formation rate as a function of 
the cosmological redshift is essential.
This function has been calculated in the framework of the 
hierarchical theory of galaxy formation (Cole et al. 1994; Kauffmann,
White \& Guiderdoni 1993; Navarro, Frenk \& White 1996).
However, in this paper we prefer to base our calculation on the star
formation rate which can be deduced from observations,
being  aware of the fact that, since these observations are reliable only 
up to redshifts of order \op z\sim 4-5,\cl
we may overlook the possibility that a big burst of star formation may
have occurred in a far past
(Rosi \& Zimmermann 1976, Bertotti \& Carr 1980, Bond \& Carr 1984). 
Following  a recently proposed approach
(Madau, Pozzetti \& Dickinson 1997), the star formation rate history that 
we have considered is based on a very simple stellar evolution model which 
assumes a time dependent star formation rate and a constant  
initial mass function. The model is directly inferred from the
impressive collection of recent observations of the rest frame 
UV-optical emission from star forming galaxies at redshifts 
ranging from zero up to $z \simeq 4-5$ 
(Steidel et al. 1996; Madau et al. 1996; Lilly et al. 1996; 
Connolly et al. 1995; Ellis 1997), and assumes  
a Salpeter initial mass function, a flat cosmology with zero 
cosmological constant and $h=0.5$. The resulting star formation
history has been extensively investigated and proves to correctly 
reproduce, within the experimental errors, the evolution of the
galaxy emission properties observed in other wavebands.
Furthermore, it  is consistent
with the indications of QSO absorption lines and metallic clouds observations
(Ellis 1997; Madau, Pozzetti \& Dickinson 1997; Lanzetta, Yahil \& 
Fernandez-Soto 1996; Pei \& Fall 1995).\\

The plan of the paper is as follows.
In Section 2 we briefly review the current state of observations
of the global star formation in field galaxies and present the model for
the star formation rate evolution that we have adopted in our
analysis; in Section 3 we
compute the  rate of collapses and the duty cycle of the process.
In Section 4 we describe the  energy
spectrum computed by Stark and Piran, which we use to model
the gravitational emission of  single events.
In Section 5 we derive the expression for the spectral energy
density and for \op \Omega_{GW},\cl and discuss the spectral properties
of this background.
In Section 6 we analyze the dependence of our results on the assumed 
model parameters. Finally, in Section 7 we present a preliminary
statistical analysis of the signal. 
A discussion of  the assumptions on which our
results are based and on  further developments of the
present work will be given in the concluding remarks.

\section{The Star Formation History}

In the past few years, our understanding of the origin and evolution of 
galaxies has greatly improved. Within the general framework of hierarchical 
clustering and gravitational instability, numerical and analytical modelling 
describe how primordial perturbations turn into 
visible stars (Cole et al. 1994; Kauffmann et al. 1994; Navarro et al. 1996).
Notwithstanding the complex physical processes at work,
the current theoretical models succeed in reproducing some of the
observed properties of the galaxy population. 
However, many fundamental aspects
still remain poorly understood (Frenk et al. 1996; Baugh et al. 1997).

The major progress made so far comes from the observational work. 
The spectacular new data obtained with the Hubble Space Telescope (HST),
Keck and other large telescopes have
extended our view of the universe up to $z\!\sim\!4\!-\!5$ 
(Steidel et al. 1996; Madau et al. 1996; Lilly et al. 1996; 
Connolly et al. 1995; Ellis 1997). 

In the following, we have attempted to summarize the relevant aspects
of the star formation history that is emerging from the data collected
so far, having in mind a reader interested in gravitational waves
but not necessarily competent in galaxy evolution.  
 
The identification of star-forming galaxies at \(2\! \stackrel
{\scriptscriptstyle <}{\scriptscriptstyle \sim}\! 
z\! \stackrel{\scriptscriptstyle
<}{\scriptscriptstyle \sim}\! 4\) in ground-based surveys and in the 
Hubble Deep Field (HDF) was successfully accomplished through the 
elaboration of robust selection criteria based on multicolor broadband 
observations of the emitter's rest-frame UV and optical stellar continuum.
Ground-based observations of star-forming galaxies at $z\approx 3$ have used 
color techniques which are sensitive to the presence of a Lyman-break 
superimposed on an
otherwise flat UV spectrum (Steidel \& Hamilton 1992). At higher redshifts,
the effect of intergalactic absorption on galaxy colors can severely affect
galaxy spectra. Its inclusion in the construction of reliable selection
criteria proved extremely efficient in 
identifying star-forming galaxies at redshifts $2\!<\!z\!<\!3.5$ (UV-dropout 
technique) and $3.5\!<\!z\!<\!4.5$ (blue-dropout technique) in the HDF images
(Madau et al. 1996). 
        
The combination of these high redshift observations with 
the recent completion of several ground-based spectroscopic surveys out to
$z\! \sim \!1$ (Lilly et al. 1996; Ellis et al. 1996) have enabled, for the
first time, a systematic study of field galaxies at increasing cosmological 
lookback times. Focusing on the entire population, a great
deal can be learned about the evolution of galaxies from the analysis 
of the integrated light emitted at each redshift in a given waveband, i.e.,
the comoving luminosity density,
\be
\rho_{\lambda}(z)=\int_0^{\infty}\!\!\!L_{\lambda}\, \phi\left(L_{\lambda},z
\right)\, dL_{\lambda},
\ee 
where $\phi\left(L_{\lambda},z\right)$ is the best-fit Schechter 
luminosity function
in each redshift bin. This quantity 
is independent of many details of galaxy evolution, such as   
the merging history or short-lived star formation episodes of individual 
galaxies, and depends on the global star formation history and on the
initial mass function (IMF) of the stars. In particular, the UV continuum
emission from an actively star-forming galaxy is mainly contributed by
massive ($M\!>\!10\,\msun$) short-lived 
($t_{MS}\!<\!2\times 10^7$yr) stars.
It has been shown that, after an initial transient phase, a steady state
is reached where the measured luminosity becomes proportional to the 
star formation rate (SFR) and independent of the galaxy history for
$t\gg t_{MS}$ (Madau, Pozzetti \& Dickinson 1997).
Assuming a universal IMF, the UV-SFR relation is determined through 
specific stellar population synthesis codes (Bruzual \& Charlot 1993; 1998)
and the global star formation rate evolution is directly
inferred from the observed UV emissivity (Madau 1997; Lilly et al. 1996).

The comoving SFR density (the mass of gas that goes into stars per unit
time and comoving volume element) as a function of redshift, 
$\dot{\rho}_*(z)$, is shown in Figure 1 for a Salpeter 
IMF and a flat cosmology ($\Omega_0=1$,
$\Lambda=0$), with $h = 0.5$. 
\begin{figure}
\begin{center}
\leavevmode
\centerline{\epsfig{figure=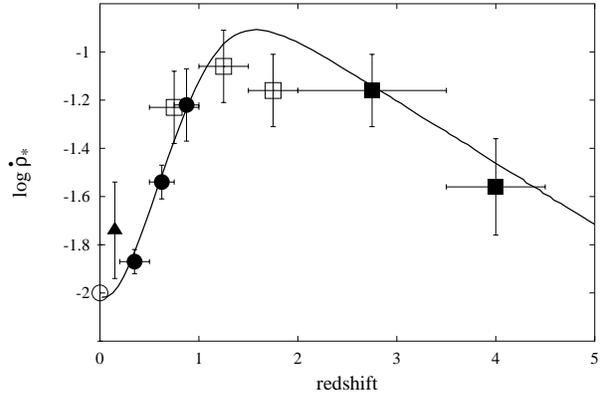,angle=270,width=8cm}}
\caption{Evolution of the log of the SFR density
($\msun \mbox{yr}^{-1} \mbox{Mpc}^{-3}$) for $\Omega_0=1$,
($\Lambda=0$), $h = 0.5$ and a Salpeter IMF. The data points are 
taken from
Gallego et al. (1995) ({\it empty dot}), Treyer et al. (1997) 
({\it filled triangle}), Lilly et al. (1996) ({\it filled dots}), 
Connolly et al. 
(1997) ({\it empty squares}) and the HDF ({\it filled squares}). The solid 
line represents the SFR density evolution assumed in the present analysis.
The fit has been kindly provided by P. Madau.}
\end{center}
\label{sfr}
\end{figure}

Despite the uncertainties due to 
the connection of disparate datasets, each using a different indicator of the
star formation activity and each selecting with a specific criterion only
some detectable subset of the overall population, the observations collected
so far appear to provide a consistent picture of the global star formation
history of the universe. The SFR density rises 
sharply from its local value to a peak at $z\!\approx\!1.5$ to fall again 
out to $z\!\approx\!4$. This evolutionary behaviour indicates that the
bulk of the stellar population was assembled in the redshift range 
$1\!\stackrel{\scriptscriptstyle <}{\scriptscriptstyle \sim}\!z\!\stackrel
{\scriptscriptstyle<}{\scriptscriptstyle \sim}\!2$ (Ellis 1997; Madau 1997). 

One major source of uncertainty is the effect of dust extinction. 
At redshift $z\sim3$,
galaxies are already enriched in heavy elements and it is likely that some
dust is mixed with gas and stars. Even a relatively small amount of dust can
attenuate the UV luminosity and reradiate the absorbed UV light in the
far-IR thus leading to an underestimation of 
the corresponding star formation activity. Dust correction factors 
are still very uncertain mainly because of the unknown shape of the dust
UV-extinction law (Pettini et al. 1997). The
amount of star formation that is hidden by dust over the entire history
of the universe will be constrained by future observations of the cosmic 
infrared background (Franceschini et al. 1997; Guiderdoni et al. 1997). 
So far, the upper
limits based on the analysis of COBE-FIRAS residuals poorly constrain 
the high-$z$ evolution of the SFR density, leaving the effect of
dust obscuration still as a matter of discussion.
    
The model for the global SFR evolution presented in Figure 1 
assumes an extinction law similar to that which applies to stars in the
Small Magellanic Cloud (SMC) and an amount of dust which results
in an upwards correction of the comoving UV luminosity by a factor $1.4$ 
at $2800$\AA\enspace and $2.1$ at $1500$\AA\enspace (Madau, Pozzetti \& 
Dickinson 1997). 

An extensive investigation of this model
has been accomplished by Madau, Pozzetti \& Dickinson (1997). 
For simplicity, the effects of cosmic chemical evolution is not 
included and all population synthesis codes assume solar metallicity, thus 
generating colors that are slightly too red for primeval galaxies
with low metallicity.
However, an encouraging similar
trend is being indicated by the gaseous and chemical evolution of the
intergalactic gas as delineated by the studies of QSO absorption lines.
The overall shape of the mean cosmological mass density contributed
by Lyman-$\alpha$ absorbers indicates that the bulk of star
formation activity should take place at $z\!<\!2\!-\!3$ (Lanzetta et al. 1996;
Storrie-Lombardi et al. 1996).

Likewise, the study of the metallic clouds in combination with chemical
evolution models leads to a volume-averaged star formation history in
agreement with the evolution inferred from galaxy surveys (Pei \& Fall 1995).
 
A  further important check on the inferred SFR history
of field galaxies comes from the study of the predicted emission properties
at UV, optical and far-infrared wavelengths. By construction, the model 
produces the right amount of ultraviolet light. While for a Salpeter
IMF the agreement with the observational data still holds
at longer wavelengths, a Scalo IMF produces too much long wavelength light
by the present epoch. Moreover, the SFR history plotted in Figure 1 appears 
able to account 
for the entire background light recorded in the galaxy counts down to 
the very faint magnitudes probed by the HDF and produces visible mass-to-light
ratios at the present epoch which are consistent with the values observed in 
nearby galaxies (Madau, Pozzetti \& Dickinson 1997).

Finally, a broad class 
of hierarchical clustering models predicts a SFR history in good agreement 
with the observationally estimated one, giving to the model a significant 
theoretical support (Baugh et al. 1997).

\section{The rate of black hole collapses}

Using the  SFR density given in Figure 1, we calculate the
evolution of the rate of core-collapse supernovae $R_{SN}(z)$,
i.e. the number of events occurring per 
unit time within the comoving volume out to redshift $z$,
\be
\label{rate1}
R_{SN}(z)=\int_{0}^{z}\!\!\!\dot{\rho}_*(z')\,\, \frac{dV}{dz'}\,\, dz'\! 
\int_{M_p}^{M_{u}}\!\!\!\! \Phi(M) \,dM .
\ee
Here $\dot{\rho}_*(z)$ is the SFR density, $dV$ the comoving
volume element and $\Phi(M)$ the IMF. The lower limit of the mass range,
$M_p$, depends on the specific nature of the supernovae considered. 
Numerical studies show that single stars with masses
$>\! 8 \msun$ evolve rather smoothly through all phases of nuclear burning
ending their life as  supernovae. For this class, 
which includes Type II and Types Ibc supernovae, the explosion occurs through
the gravitational collapse of their core, leaving behind a neutron 
star or a black hole (e.g. Ruiz-Lapuente 1997). The smallest progenitor mass 
which is expected to lead to a black hole ranges from $18 \msun$ to $30\msun$, 
depending on the iron core masses expected from stellar evolution calculations
(Timmes, Woosley \& Weaver 1995; Woosley \& Timmes 1996). 
In the calculations to follow we will adopt a reference value of 
$25 \msun$ and 
we will discuss the implications of higher values of $M_p$ on our results
in Section 6.

As previously mentioned, \op \dot{\rho}_*(z)\cl has been calculated by
assuming a Salpeter IMF, which we will accordingly use for our
calculations,
\be
\label{salpeter}
\Phi(M) \propto M^{-(1+x)}, \,\,\,\,\, \mbox{with} \,\,\,x=1.35,
\ee
normalized through the relation
\be
\int_{M_{l}}^{M_{u}}\!\! \! M \,\Phi(M)\, dM = 1,
\ee
with $M_{l}=0.1\msun$ and $M_{u}=125\msun$. 

As recently suggested by Madau (1998), 
the evaluation of the rate of core-collapse supernovae based on the SFR
density previously discussed and on an assumed universal IMF
is largely independent of the choice of the IMF.
Using an extreme approach, we may assert that we do not need an IMF 
to deduce the rate of supernovae explosion. In fact, the SFR density is 
inferred from the rest-frame
UV continuum emission, and the stars which are responsible for this emission
are the more massive ones, which, in turn, are those that at the end of
their evolution give origin to core-collapse supernovae.

The model we use for the SFR history assumes a flat cosmology ($\Omega_0=1$) 
with vanishing cosmological constant and $h = 0.5$, which will be the 
values of the cosmological parameters we will adopt in the rest of this paper.
In this case, the comoving volume element is related to $z$ through,
\be
\label{volcom}
\frac{dV}{dz}\!= \!16 \pi \left(\frac{c}{H_0}\right)^3
\!\frac{[1-(1+z)^{-1/2}]^2}{(1+z)^{3/2}}, 
\ee
We have evaluated the integral (\ref{rate1}) by using the star formation
rate density \op \dot{\rho}_*(z)\cl plotted in Figure 1,
and eqs. (\ref{salpeter}) and (\ref{volcom}),
for two values of \op M_p.\cl  The lowest value of the mass cutoff
\op M_p=8\msun\cl corresponds to the total rate of core-collapses 
which produce either neutron stars or black holes. 
$ M_p=25\msun\cl is a possible  mass cutoff for the
production of black holes at the present state of our knowledge
(Woosley \& Timmes 1996).
The results are plotted in Figure 2.
\begin{figure}
\begin{center}
\leavevmode
\centerline{\epsfig{figure=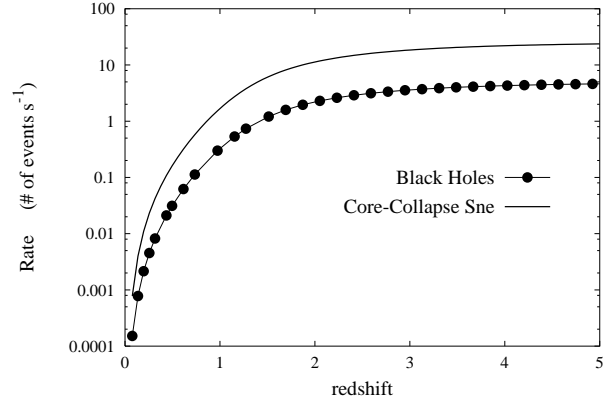,angle=270,width=8cm}}
\caption{Evolution of the rate of supernovae as a function of redshift.
The upper curve represents the rate of core-collapse supernovae (Type II)
and assumes a lower mass cut-off of $M_p=8 \,\mbox{M}_{\odot}$. 
The lower curve represents only those events that lead to 
black hole collapses and assumes $M_p=25 \,\mbox{M}_{\odot}$.}
\end{center}
\label{rate}
\end{figure}

The predictions for the rate of supernovae are in good agreement
with the observed local values (Cappellaro et al. 1997).
At higher redshift, $z\!\stackrel{\scriptscriptstyle >}
{\scriptscriptstyle \sim}1$, the detection of Type II events
must await future experiments, such as the Next Generation Space Telescope
(NGST) (Madau, Della Valle, Panagia 1998; Sadat et al. 1997).
It should be noted that the rate shown in 
Figure 2, which is integrated over the comoving volume as 
indicated in eq. (\ref{rate1}),
tends to saturate to a constant value: this 
follows from the drop of the SFR density at high redshift.
Conversely, the rate per unit comoving volume, would trace the 
rise and drop of the SFR density evolution.

Finally, the total number of supernovae explosions per unit time 
leading to black hole formation is obtained by integrating  eq. (\ref{rate1})
for \op z=\infty,\cl  
\be
\label{totalrate}
R_{BH}=
4.74 \,\,\mbox{events/s} \qquad\hbox{for}\qquad M_p=25\, \msun.
\ee
A significant quantity which indicates whether the collective effect of the
bursts of gravitational waves emitted in such collapses  generates a
continuous background is the duty cycle. It is defined as
\be
\label{duty}
D(z)=\int_0^z dR_{BH}(z) \,\overline{\Delta\tau}_{GW}\, (1+z) ,
\ee
where $D(z)$ is to be interpreted as the duty cycle now of all bursts
generated back to redshift $z$ and  
 \op  \overline{\Delta\tau}_{GW}  \cl is the average time duration of 
single bursts at the emission, which we will assume to be 
\op\sim 1\,\mbox{ms}.\cl
The total duty cycle is
\be
\label{dutytotal}
D= 1.57 \times 10^{-2}\qquad\hbox{for}\qquad M_p=25\msun.
\ee
The smallness of the duty cycle implies that our stochastic gravitational-wave 
background cannot be considered as a continuous one but rather as a 
{\em shot-noise} process, consisting in a sequence 
of widely spaced bursts with mean time separation of $\sim \! 0.2$ s much 
larger than the typical duration of each single burst, which is of order a few 
ms. 

It has recently been suggested that supernovae occurring soon
after galaxy formation could contribute a duty cycle of order
unity, creating a nearly continuous background (Blair \& Ju 1997).
These investigations are based on a supernova rate obtained 
from its observationally estimated local value through an integration 
of the source count equation. Therefore, the important effect of the global 
SFR evolution was not included, leading to an overestimation of the 
rate. 

The implication of the evolving galaxy population
has been investigated by Kosenko \& Postnov (1998) to 
estimate the stochastic gravitational background produced 
by extra-galactic merging binary white dwarfs.
They show that when the global SFR evolution is included, 
the level of this extra-galactic background
can be comparable with the corresponding galactic background signal. 
 
\section{The energy spectrum of a core-collapse to a black hole}

The energy spectra that we shall use to model the  process of 
collapse to a  black hole
have been obtained  by Stark \& Piran (1985, 1986).
They numerically  integrated  the fully non-linear
system of Einstein + hydrodynamics equations describing the 
collapse  of a star with a polytropic equation of state
\op p= K\rho^\Gamma,\cl and adiabatic index \op\Gamma=2. \cl
The initial configuration  is a  spherically
symmetric very compact star 
with central density \op 1.9 \times 10^{15}\,(M_{core}/\msun)^{-2}\cl 
g cm$^{-3}$ and radius 
\op 8.8\times 10^5 \, (M_{core}/\msun)\, \mbox{cm}\cl
which, as a consequence of a sudden
reduction to a fraction $f_p$ (with $f_p=0.01$ or 0.4) of the equilibrium 
central pressure, starts to collapse.
Simultaneously,  the star is given an angular momentum
distribution, approximating rigid-body rotation. 

Thus, the collapse which is considered is that
of  a compact naked core. 
The studies on stellar evolution show that compact cores
(typically degenerate iron cores) which  form in the interior of  massive
stars, are surrounded by layers of lighter material.
The dynamics of this envelope is not considered in the simulation of
Stark and Piran. 
In the concluding remarks we will return on this point,
which,  however, is not of substantial relevance for our present
calculations.

Stark and Piran find that if the total rotation parameter 
\op a \equiv J/\left(GM_{core}^2/c\right)\cl exceeds a
critical value 
\beq
a_{crit}&=&1.2\pm 0.2\qquad\qquad\hbox{if}\qquad f_p=0.01,\\
\nn
a_{crit}&=&0.80\pm 0.05\qquad \,\,\,\,\,\,\hbox{if}\qquad f_p=0.4,
\eeq
the rotational energy dominates, the star
bounces and no collapse occurs.
Conversely, if \op a < a_{crit}\cl a black hole forms, and
a strong  burst of gravitational waves  is emitted, with the following
characteristics.
\begin{figure}
\begin{center}
\leavevmode
\centerline{\epsfig{figure=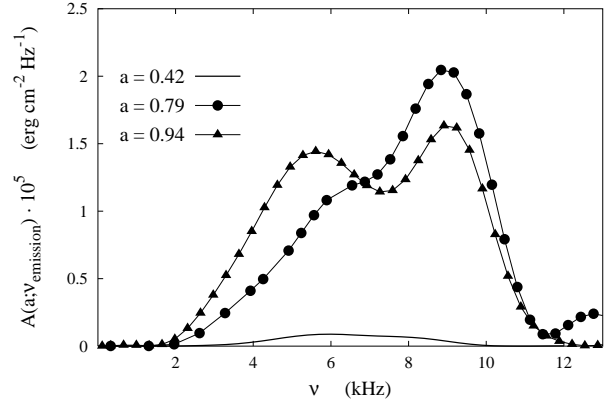,angle=270,width=8cm}}
\caption{The average energy flux emitted during the axisymmetric collapse
of a rotating, polytropic star to a black hole of $M_{core}=1.5 \, \msun$ at
a distance of $15$ \, Mpc. The three curves correspond to assigned values of 
the rotational parameter and refer to \op f_p=0.01$.}
\end{center}
\label{starkfig}
\end{figure}

Let \op h^{TT}_{\theta\theta}(t,r,\theta,\phi)\cl 
and \op h^{TT}_{\theta\phi}(t,r,\theta,\phi)\cl
be the physical components of the radiative part of the metric
tensor evaluated in the transverse-traceless gauge.
The energy flux radiated per unit time
by the collapsing source is given by
the \op (01)$-component of the stress-energy tensor of gravitational 
waves, 
\beq
\label{stress}
T_{01}= \frac{dE}{dS dt} =
{1\over {16\pi }}\left\{ \left[ \dot h^{TT}_{\theta\theta}
(t,r,\phi,\theta)\right]^2\!\!+\!\!
\left[ \dot h^{TT}_{\theta\phi}(t,r,\phi,\theta)\right]^2\right\} 
\eeq
(in geometric units, $c=G=1$), where \op x^0=t, x^1=r, x^2=\phi, x^3=\theta.\cl
If we Fourier transform, 
\be
\label{fouriertransf}
h^{TT}_{\mu\nu}(\omega,r,\theta,\phi)=
\int_{-\infty}^{+\infty}\!\!h^{TT}_{\mu\nu}(t,r,\theta,\phi)
\,e^{-i\omega t}\,dt,
\ee
and apply Parseval's theorem, the energy flux per unit frequency
can be defined as
\be
\label{dedome}
\frac{dE}{dS d\omega}={\omega^2\over {32\pi^2}}\left\{
\vert h^{TT}_{\theta\theta} (\omega,r,\phi,\theta)\vert^2+
\vert  h^{TT}_{\theta\phi}(\omega,r,\phi,\theta)\vert^2
\right\},
\ee
and the energy flux per unit  solid angle, 
\be
\label{dedomesolid}
\frac{dE}{d\Omega d\omega}={\omega^2 r^2 \over {32\pi^2}}\left\{
\vert h^{TT}_{\theta\theta} (\omega,r,\phi,\theta)\vert^2+
\vert  h^{TT}_{\theta\phi}(\omega,r,\phi,\theta)\vert^2
\right\}.
\ee
For  \op\omega \geq 0,\cl eq. (\ref{dedome}) and  eq. (\ref{dedomesolid}) 
must be multiplied by a factor of 2 in order to account
for the folding of negative frequencies into positive.
We finally define an average energy flux per unit frequency,
\op f(\nu),\cl
by integrating eq. (\ref{dedomesolid}) over the solid
angle, and dividing by \op 4\pi r^2 \cl
\be
\label{fdinu}
f(\nu)\equiv
\frac{1}{4\pi r^2}\int_0^{2\pi}\!\!d\phi\int_0^{\pi}
\left(\frac{dE}{d\Omega d\nu}\right)\, \sin\theta\, d\theta.
\ee
To express \op f(\nu)\cl in physical units, eq. (\ref{fdinu})
must be multiplied by the factor \op c^3/G.\cl

Let us now consider the collapses studied by Stark and Piran
that belong to the case \op f_p=0.01.\cl
During an axisymmetric collapse, the energy is emitted in both
polarization modes but since \op h_+ \approx 10 \, h_{\times},\cl we shall
consider only the contribution of the \op h_+ \cl component. 
The function \op f(\nu)\cl 
is plotted  in Figure 3
for a black hole of \op M_{core}=1.5\msun\cl 
which forms in the Virgo cluster, for assigned values of the angular
momentum.
The quantity 
$f(\nu)\cl attains a sharp maximum for some value of
the frequency which depends on the angular momentum; for instance, we find 
$\nu_{max}=8.6$ kHz for $a=0.79$ and $\nu_{max}=9.2$ kHz for $a=0.94$. 

It is interesting to compare the location of this maximum with
the frequency of the lowest \op m=0\cl quasi-normal mode of
a rotating black hole with the same mass and angular momentum 
(e.g. Leaver 1986). For \op\ell=2\cl we find
$\nu_{0}=8.6$ kHz, for $a=0.79$, and 
$\nu_{0}=9.0$ kHz, for $a=0.94$; 
for \op a=0\cl it would be \op\nu_0\sim 8$ kHz. 
Thus the peak of \op f(\nu)\cl is located at a
frequency which is very close to the frequency of the
lowest \op m=0\cl quasi-normal mode.
This means that a substantial fraction of  energy
will be emitted after the black hole has formed: 
it will oscillate in its quasi-normal modes 
until its residual mechanical energy is  radiated away in gravitational
waves.

For values of the rotation parameter in the range 
\op  0.2 < a < 0.8,\cl the amplitude of
the peak of \op f(\nu)\cl scales as \op a^4.\cl
For high values of the angular momentum,
the  star becomes flattened into the equatorial plane
and then bounces vertically, but still continues to collapse
inward until the black hole is formed. In this case
the amplitude of \op f(\nu)\cl in the low frequency
region increases, and a further peak appears,
the amplitude of which may become comparable to that of
the peak corresponding to the quasi-normal modes.

According to Stark and Piran 
the main characteristics of the energy spectra and waveforms 
for \op f_p=0.01\cl and for \op f_p=0.4\cl are similar.
If \op f_p=0.01\cl the star becomes more flattened and the
collapse is slightly faster. The ratio between the maximum of
the energy spectrum for \op f_p=0.01\cl and for \op f_p=0.4,\cl
for angular momenta close to the critical value, is
\op \sim 5.\cl

In general, the efficiency of the process 
of {\it axisymmetric} core-collapse to a black hole
studied by Stark and Piran is 
$\Delta E_{GW}/M_{core} c^2 \leq 7 \times 10^{-4}$. 
It should be remembered that less symmetric situations may result in a more
efficient production of gravitational waves.

\section{Spectral properties of the stochastic background}

The spectral energy density
of the stochastic background produced by those gravitational 
collapses that led to black holes formation,
can be obtained by integrating over the allowed range of masses and redshifts
the differential rate of black hole formation,
\op dR_{BH}(M,z),\cl
multiplied by the  energy spectrum of a single event.\footnote{The 
smallness of the total duty cycle implies that, on average, different 
bursts do not overlap, so that interference among the waves generated by 
different collapses is unlikely to occur.} 
Thus, we need to know how the energy flux produced by
a star of mass \op M\cl and angular momentum \op a,\cl which collapses at a
given redshift \op z,\cl would be observed today.

The average energy flux per unit frequency in the source rest frame,
which is plotted in Figure 3
for a core of \op 1.5\,\msun\cl at the distance of \op 15\, \mbox{Mpc}\cl
can be written as follows
\be
f(\tilde\nu)=  A(a,\tilde\nu)\,
\left(\frac{M_{core}}{1.5 \mbox{M}_{\odot}}\right)^2
\left(\frac{d}{15\mbox{Mpc}}\right)^{-2},
\ee
where  \op d\cl is the distance from the source in the source rest frame,
and 
\be
\tilde\nu=\nu_{emission}\,\left(\frac{M_{core}}{1.5\msun} \right).
\ee
Such a signal, if observed here and now, would give
\be
\label{flux}
f(\nu_{obs})=A(a,\tilde\nu)\,
\left(\frac{M_{core}}{1.5 \mbox{M}_{\odot}}\right)^2
\left(\frac{d_L(z)}{15\mbox{Mpc}}\right)^{-2},
\ee
where  the observational frequency \op \nu_{obs}\cl is related to
\op \tilde\nu\cl through
\be
\label{nuobs}
\nu_{obs}=\frac{\tilde\nu}{(1+z)}\, \frac{1.5\msun}{M_{core}}
\ee
and 
\be
d_L(z) = \frac{2c}{H_0}\,(1+z)[1-(1+z)^{-1/2}]
\ee
is the luminosity distance. 
In the following we shall assume that the fraction of the progenitor star
which collapses is $M_{core}=\alpha~M$ with \op \alpha=0.1.\cl

The differential rate of black hole collapse is given by
[see eq. (\ref{rate1})]
\be
\label{diffrate}
dR_{BH}(M,z)=\dot{\rho}_*(z)\, \frac{dV}{dz} \,\Phi(M)\, dM \,dz.
\ee
Thus the spectral energy density is
\be
\label{eq:ltot}
\frac{dE}{dt dS d\nu}= \int_0^{\infty}\!\!\int_{M_p}^{M_u}\!\! f(\nu_{obs})\,
dR_{BH}(M,z).
\ee
Since we do not know the distribution of angular momenta 
of the formed black holes, we evaluate the spectral energy density
for assigned values of the rotation parameter. 
We consider three possible values of $a$, namely 
$a=0.42,0.79,0.94$, 
and  give the corresponding functions \op A(a,\tilde{\nu})\op as 
an input to the numerical code.
For each chosen value of $a$,  we solve the integral
(\ref{eq:ltot})  as a function of  the
observation frequency \op \nu_{obs}\cl.

The statistical description of a stochastic gravitational background
relies on three basic assumptions, i.e. that the signal is isotropic,
stationary and unpolarized (Allen \& Romano 1998, Maggiore 1998).
In our present case, the background is produced by
a large number of independent gravity-wave sources extending 
over the redshift range $0<z<4-5$. Thus, the first two assumptions are
certainly satisfied and,
assuming that the sources have random orientations, the radiation incident
on the detector has statistically equivalent polarization components
(Thorne 1987). 
From the spectral energy density we shall evaluate the corresponding 
values of the spectral strain amplitude,
\be
\label{strain}
\sqrt{S_h(a,\nu_{obs})}=\left(\frac{2 G}{\pi c^3} 
\frac{1}{\nu_{obs}^2}\right)^{1/2}\left(\frac{dE}{dt dS d\nu}\right)^{1/2}, 
\ee
and of the closure energy density of gravitational waves, 
\be
\Omega_{GW}(a,\nu_{obs})= \frac{\nu_{obs}}{c^3 \rho_{cr}}\, \frac{dE}{dt dS
d\nu}
=\frac{4}{3}\frac{\pi^2}{H_0^2}\,\nu_{obs}^3 \, S_h(a,\nu).
\ee
The results are plotted in Figures 4, 5 and 6.
\begin{figure}
\begin{center}
\leavevmode
\centerline{\epsfig{figure=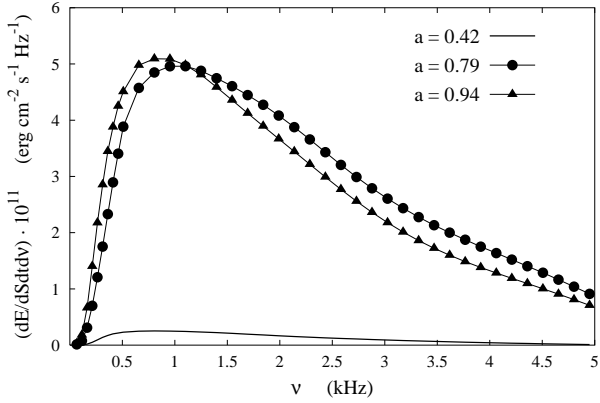,angle=270,width=8cm}}
\caption{The spectral energy density \op (dE/dSdtd\nu)\cl
is plotted  as a function of the observational frequency.
The three curves correspond
to assigned values of the rotational parameter. }
\end{center}
\end{figure}
\begin{figure}
\begin{center}
\leavevmode
\centerline{\epsfig{figure=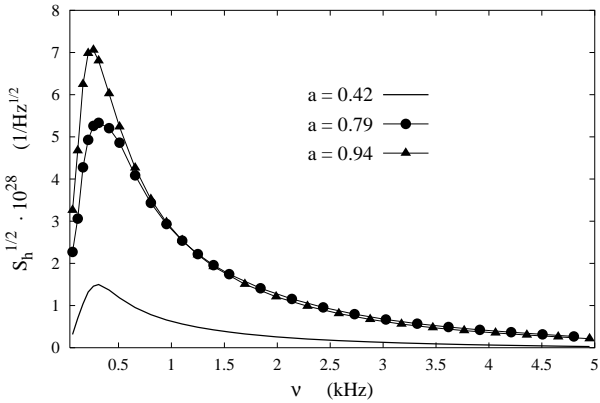,angle=270,width=8cm}}
\caption{The spectral strain amplitude \op S_h^{1/2}\cl corresponding to the
spectral energy density given in Figure 4 is plotted  as a function of the
observational frequency.}
\end{center}
\label{sh}
\end{figure}
\begin{figure}
\begin{center}
\leavevmode
\centerline{\epsfig{figure=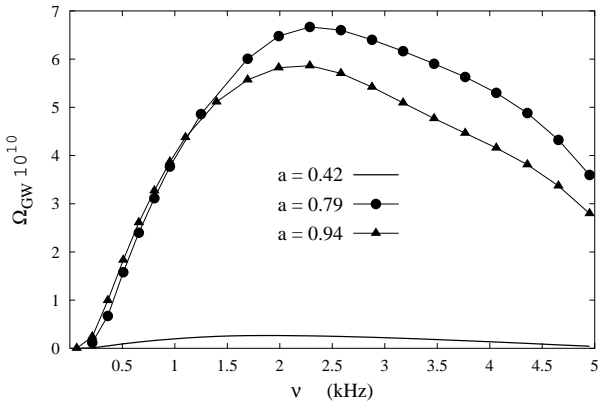,angle=270,width=8cm}}
\caption{The function \op\Omega_{GW}\cl
corresponding to the spectral energy density plotted in Figure 4.}
\end{center}
\label{omega}
\end{figure}

\section{Dependence on the assumed model parameters}

Let us now discuss how the spectral energy density 
and  the spectral strain amplitude computed in Section 5
depend on some of the parameters we
use in our calculations. First of all the mass cutoffs.

The lower cutoff on the mass of the progenitor star whose collapse
produces a  black hole, and  which we set at
$M_p=25 \msun$,  depends on the stellar evolution model upon which the 
calculations describing the life of massive stars are based,
as well as on the effect of fallback during the explosion (Woosley \& Timmes
1996 and reference therein). 
Thus, we have investigated the consequences
that a higher value of this cutoff, namely $M_p=30 \, \msun$, 
would introduce in our results and  
we have similarly considered the effect of reducing the dynamical range
lowering the upper mass cutoff from \op 125 \, \msun\cl to $60 \, \msun$.
The  various dynamical ranges considered are listed in
Table 1, where we also give the values of the 
rate of collapses to a black hole and the duty cycle.
\begin{table*}
\caption{The different combinations of low and high mass
cutoffs considered in the analysis of the dependence of the results
on the assumed model parameters.}
\begin{tabular}{@{}lllll@{}}
Model & $M_p/\msun$ & $M_{u}/\msun$& $R_{BH}$(events/s) 
& $D$\\\hline
      &     &   &   &\\[3pt]
case A & 25 & 125 & 4.74 & $1.57 \times 10^{-2}$\\
case B & 30 & 125 & 3.57 & $1.19 \times 10^{-2}$\\
case C & 25 & 60  & 3.71 & $1.23 \times 10^{-2}$
\end{tabular}
\end{table*}
We see that if the lower mass cutoff is raised to 
$M_p=30\,\msun$ the values of the total rate and duty cycle
are a factor $\sim 1.33$ lower than in the $M_p=25 \,\msun$ case.
The effect of lowering the upper mass cutoff to $60\, \msun$ is less 
significant. 
Infact, although the percentage of progenitors having mass between 
$60$ and $125 \msun$ is about $38\%$, when we compute 
the rate as in eq. (\ref{rate1}), since the IMF goes
like \op \sim M^{-2.35},\cl bigger masses will give
smaller contributions to the integral, and 
the change in rate and duty cycle will be small.

As for the spectral energy density, since the frequencies of
the energy spectrum emitted during a single collapse 
are inversely proportional to the mass of the collapsing core 
(which we have assumed to be roughly 10 \% of the progenitor mass),
the major effect of shifting  the dynamical range towards 
higher values of the progenitor mass will be to 
reduce the power at high frequency and
decrease the overall amplitude. 
Similarly, the effect of reducing the 
upper mass cutoff is that of loosing  some power at low frequencies.
In addition, from Figure 7  we see that the effect of increasing \op M_p\cl
to \op 30\,\msun\cl is less significant than that of lowering \op M_u\cl
to \op 60\,\msun.\cl The reason is immediately understood if we look at
eq. (\ref{eq:ltot}) from which \op dE/dt dS d\nu\cl 
is computed.
It involves an integral over the mass which, due to the
explicit  dependence of
\op f(\nu_{obs})\cl and of \op \Phi(M)\cl on \op M$, goes like \op M^{-0.35}$.

As for the spectral strain amplitude, the \op 1/\nu_{obs}\cl factor 
in eq. (\ref{strain}) 
introduces a factor \op M,\cl through eq. (\ref{nuobs}), which makes 
negligible the change in \op \sqrt{S_h}\cl when the lower mass cutoff is 
raised to \op 30~\msun\cl. Conversely,
the  effect of changing the upper mass cutoff is
to decrease the amplitude of the peak by almost a factor of 2. 

The results of this discussion are plotted in Figures 7 and 8 for
a single value of the rotational parameter $a=0.79$ and the for various 
combinations of the mass ranges given in Table 1.
In Table 2 we give the frequency at which the maximum 
of \op \left(dE/dt dS d\nu \right)\cl and \op \sqrt{S_h}\cl
is located in the various cases, and the corresponding amplitudes.

It is also interesting to see how the strain amplitude changes
if we choose different values of the parameter \op\alpha,\cl i.e.
of the fraction of the progenitor star which collapses. 
If \op \alpha\cl increases, the mass of the collapsing core increases.
As a consequence, the emitted energy, which is proportional of \op
M_{core}^2\cl, increases, and the frequency of the maximum, which is
inversely proportional to \op M_{core}\cl decreases.
This results in an overall higher amplitude of \op S_h^{1/2},\cl and in 
the moving of the peak frequency toward smaller values.
The results are shown in Figure 9.

\begin{table*}

\caption{The values of the frequencies at 
which \op dE/dt dS d\nu \cl  
and \op S_h^{1/2}\cl have a maximum,
and the corresponding amplitude
(respectively in erg cm$^{-2}$ s$^{-1}$ Hz$^{-1}$ and in Hz$^{-1/2}$)
are tabulated for different combinations
of the upper and lower cutoff parameters.}
\begin{tabular}{@{}lrrrrrl@{}}
Model & $M_p/\msun$ & $M_{up}/\msun$& $\nu_{max}$(\mbox{Hz})& 
$\left(\frac{dE}{dSdtd\nu}\right)_{max}$ &
$\nu_{max}(\mbox{Hz})$& $\sqrt{S_{h~max}}~(\mbox{Hz}^{-1/2})$\\ \hline
      &     &      & \\[3pt]
case A & 25 &   125 &$1.05 \times 10^3$ &$5.0 \times 10^{-11}$ &
$3.12 \times 10^2$ & $5.3 \times 10^{-28}$\\
case B & 30 & 125 &$1.05 \times 10^3$ & $4.9 \times 10^{-11}$ &
$3.12 \times 10^2$ & $5.4 \times 10^{-28}$\\
case C & 25 & 60 & $2.55 \times 10^3$& $4.0 \times 10^{-11}$ &
$7.56 \times 10^2$ & $2.4 \times 10^{-28}$
\end{tabular}
\end{table*}

\begin{figure}
\begin{center}
\leavevmode
\centerline{\epsfig{figure=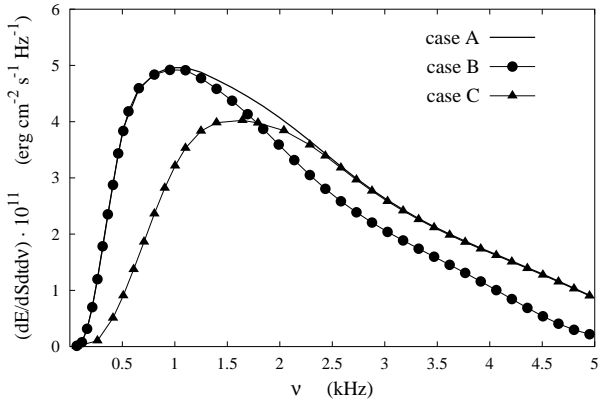,angle=270,width=8cm}}
\caption{The spectral energy density \op dE/dSdtd\nu \cl 
is plotted for various combinations of the mass cutoffs (see Table 1).
The curves correspond to a unique value of the rotational parameter
$a=0.79.$}
\end{center}
\label{vmspectra}
\end{figure} 

\begin{figure}
\begin{center}
\leavevmode
\centerline{\epsfig{figure=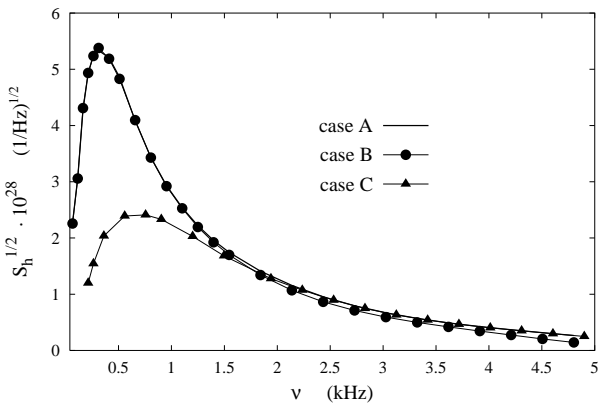,angle=270,width=8cm}}
\caption{The spectral strain amplitude \op S_h^{1/2}\cl 
corresponding to the spectral energy density given in Figure 6.}
\end{center}
\label{vmsh}
\end{figure}

We have finally analysed the consequences of a different evolution
history  of
the SFR density at high redshift, assuming a constant tail at
$z>1-2$ at approximately the amplitude of the maximum.
This evolutionary behaviour is still consistent with the
upper limits found in the analysis of COBE-FIRAS residuals
(once the black body contribution of the cosmic microwave background has been
subtracted off) in the infrared and is predicted by 
models of galaxy evolution in which dust obscuration significantly
attenuates the UV emission at high redshifts (Franceschini et al. 1997).
The overall effect of an increased SFR at high redshifts is negligible 
on  the spectral energy density $dE/dSdtd\nu$ as well as on
 $\Omega_{GW}$ and sligthly increases the amplitude of $\sqrt{S_h}$ 
at low frequencies (see Figure 10). 

Thus, the results of our analysis are not seriously affected by 
the uncertain amount of dust extinction, and the possibility that 
UV-optical data may lead to an underestimation of the global SFR at high
redshifts does not invalidate our main conclusions.
The reason for the negligible role played by high redshift sources
is that the average energy flux 
of gravitational waves contributed by each source decreases as the inverse 
of the squared luminosity distance. 
  
\begin{figure}
\begin{center}
\leavevmode
\centerline{\epsfig{figure=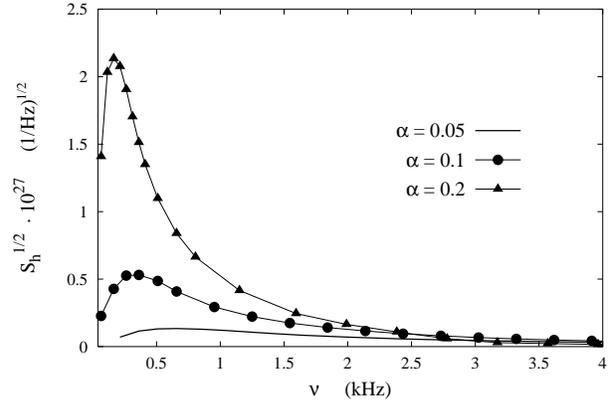,angle=270,width=8cm}}
\caption{The spectral strain amplitude \op S_h^{1/2}\cl is plotted
for three  different values of the parameter \op\alpha\cl which 
defines the fraction of the progenitor star which collapses.
The curves correspond to $a=0.79$.}
\end{center}
\end{figure}
 
\begin{figure}
\begin{center}
\leavevmode
\centerline{\epsfig{figure=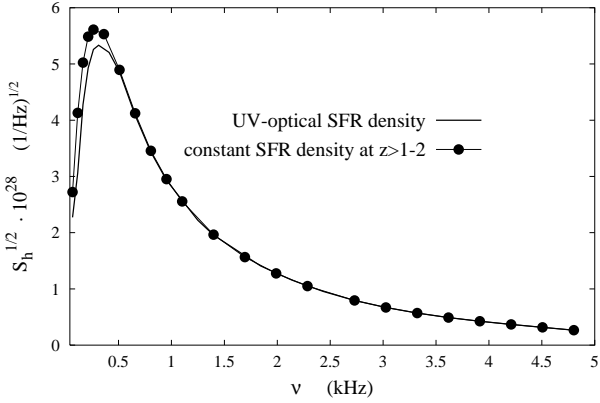,angle=270,width=8cm}}
\caption{The spectral strain amplitude \op S_h^{1/2}\cl is plotted
for two different evolutionary models for the SFR density:
the continuous line refers to the SFR shown in Figure 1,
the dotted line uses the same SFR but with a constant tail
at \op z > 2.\cl (see text). 
The curves correspond to a unique value of the rotational parameter
$a=0.79$.}
\end{center}
\end{figure}

\section{Statistical properties of the stochastic background}

In Sections 5 and 6, the stochastic background produced by 
gravitational collapses to black holes throughout the universe 
was considered to be continuous. This is indeed the case as long as
an observational run is much longer than the mean temporal interval between 
two successive bursts. This is given by the inverse of the total rate of
events and can be estimated to be $\simeq 0.2$ s (see Section 3).
Likewise primordial stochastic backgrounds, the statistical properties of 
the signal will be similar to the detector noise but with lower amplitude and
the optimal detection strategy will consist in performing a 
correlation between at least two detectors (Allen 1996; Allen \& Romano 1997).  

As an alternative strategy, one could try to exploit the characteristic
shot-noise structure of the signal and design a specific algorithm to
extract it from a single detector noise. We plan to discuss this possibility
in a forthcoming paper. As a preliminary step, we
shall now discuss some statistical properties of this background.
We have computed the number of individual signals that
are expected to have a maximum  amplitude $h_{max}$
above a given threshold, $h_t$,  within an observational run $\Delta T_{obs}$
\be
{\cal N}_{\Delta T_{obs}}(h_{max}\geq h_t) = {\cal P}(h_{max}\geq h_t) 
\,\Delta T_{obs}\, R_{BH},
\ee
where ${\cal P}(h_{max}\geq h_t)$ is the probability that $h_{max}$ is 
above the threshold, and $R_{BH}$ is the total rate of collapses to a
black hole.

Following Stark and Piran, the expression of the maximum 
amplitude  of the gravitational wave  emitted in a 
single black hole collapse  can be written as
\be
h_{max}(M,z)
=C(a) \, \frac{\alpha\, M}{1.5 \,\msun}\,\left(\frac{d_L(z)}{15 \mbox{Mpc}}
\right)^{-1}
\ee
where $C(a)$ is a constant term which is related to the rotational parameter
$a$ through,
\[C(a)=\left\{\begin{array}{ll}
                4.79 \times 10^{-22} \, a^2 & \mbox{if $a<0.8$} \\
                2.86 \times 10^{-22}        & \mbox{otherwise}.
            \end{array}
\right. \]   
Thus, $h_{max}(M,z)$ depends on the mass of the progenitor star and on the 
redshift at which  the collapse occurs. 
Therefore, its probability distribution can be 
obtained from the probability distributions of the mass and redshift
$p(M,z)$ by performing the following integral,
\be
p(h) = \int\! dM\,dz\,\, p(M,z)\,\, \delta_D(h_{max}(M,z)-h),
\ee 
where $h$ is any value of the random variable $h_{max}$ and $\delta_D$ is the
Dirac delta function. 

The function \op p(M,z)\cl is computed in the following way. The number of
black hole collapses with progenitor mass in the range \op (M,M+dM)\cl
occurring at a redshift between \op (z,z+dz)\cl is,
\be
d{\cal N} = \Phi(M) \,dM\,\Psi(z) \,\frac{dV}{dz} \,dz.
\ee
Here \op \Psi(z)\cl is the mass of gas that goes into stars per unit comoving
volume, i.e.,
\be
\Psi(z)=\int_{0}^{z}\!\! \dot{\rho}_{*}(z') \,
\left|\frac{dt}{dz'}\right|\, dz',
\ee
where, in our case the cosmic time is related to \op z\cl through 
$dt/dz= - H_0^{-1}\,(1+z)^{-5/2}$.
Thus, the probability that a black hole collapse occurs with progenitor mass 
in the range \op (M,M+dM)\cl and redshift in the range 
\op (z,z+dz)\cl is,
\be
p(M,z)\, dM \,dz =\frac{d {\cal N}}{{\cal N}_{tot}} 
\ee
where \op {\cal N}_{tot}\cl is the total number of events. 
The probability of a signal with \op h_{max}\geq h_t,\cl  
\be
{\cal P}(h_{max}\geq h_t)=\int_{h_t}^{\infty}\!\!dh\,\, p(h)
\ee
is plotted in Figure 11 as a function of
$h_t$ for the same values of the rotational parameter considered in
Section 5.\\

\begin{figure}
\begin{center}
\leavevmode
\centerline{\epsfig{figure=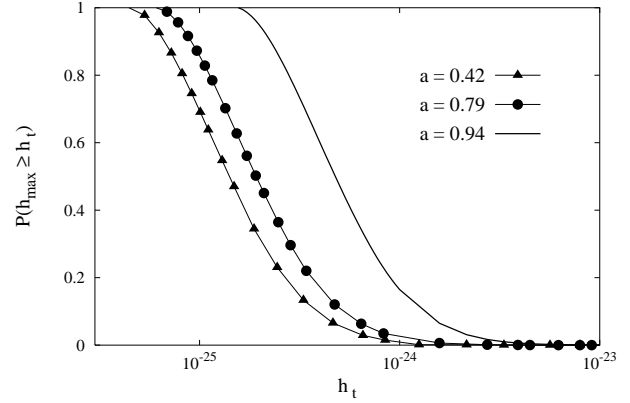,angle=270,width=8cm}}
\caption{The probability of detecting a signal with maximum observable strain 
amplitude $h_{max}$ greater than a threshold value $h_t$,
is plotted as a function of 
$h_t$ for assigned values of the rotational parameter.}
\end{center}
\label{probability}
\end{figure}
In Figures 12 and 13 we show the corresponding 
number of events expected \op {\cal N}_{\Delta T_{obs}}(h_{max}\!\geq\! h_t)
\cl for an observational run of 1 s and 1 yr respectively.

\begin{figure}
\begin{center}
\leavevmode
\centerline{\epsfig{figure=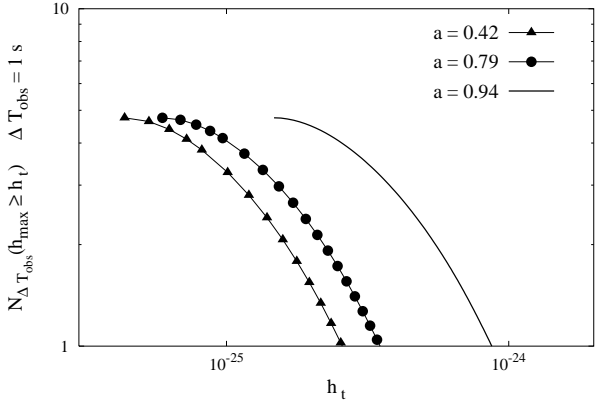,angle=270,width=8cm}}
\caption{The number of events with $h_{max}>h_t$ in an observational run
of 1 sec as a function of $h_t$ 
for three assigned values of the rotational parameter.}
\end{center}
\label{numbersec}
\end{figure}
\begin{figure}
\begin{center}
\leavevmode
\centerline{\epsfig{figure=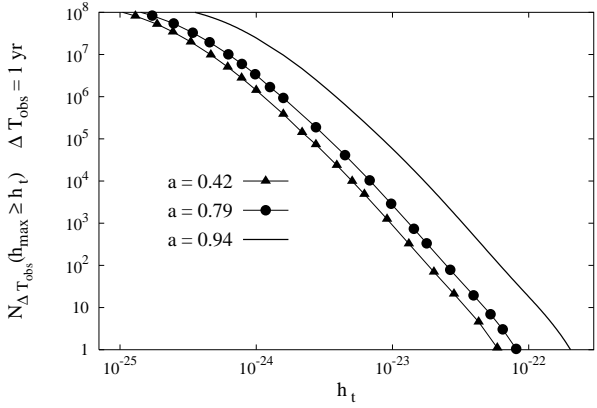,angle=270,width=8cm}}
\caption{The number of events with $h_{max}>h_t$ in an observational run
of 1 yr as a function of $h_t$ 
for three assigned values of the rotational parameter.}
\end{center}
\label{numberyr}
\end{figure}

\section{Concluding Remarks}

From the analysis carried out in this paper it emerges 
that a reliable estimate of the 
stochastic background of gravitational waves produced by
a cosmological population of astrophysical sources 
cannot set the important effect of SFR evolution aside.

The extraordinary advances attained in observational cosmology 
are leading to a coherent picture for the global star formation history
in field galaxies that can be used to infer the redshift evolution of
the rate of gravitational wave sources.   

More importantly, the resulting gravitational signal
appears to be insensitive to the uncertainties that still trouble 
the high redshift SFR observations (as a consequence of the uncertain amount
of dust obscuration), as the main contribution to the 
resulting $\Omega_{GW}$ and $\sqrt{S_h}$ comes from low-to-intermediate 
redshift sources. 

We have restricted our investigation to a specific cosmological
background model with $\Omega_0=1, \Lambda=0, H_0=50\,\mbox{km}\,\mbox{s}^{-1}
\mbox{Mpc}^{-3}$. We plan to extend the analysis to different choices 
of background cosmology in a subsequent paper. We can anticipate that, 
if we assume that the cosmological constant is zero,
the SFR density decreases with $\Omega_0$. At the same time, the geometrical 
effect on the comoving volume element increases the rate of black hole 
collapses compared to a flat background, although the
decrease of the single event spectra with the square of the inverse 
luminosity distance partly cancels the effect. Altogether we would then expect 
a modest increase of amplitude of the resulting gravity-wave background. 

The spectral properties of the background of gravitational waves 
discussed in this paper depend on the energy spectrum  we have chosen as
representative of the process of gravitational collapse to a black hole.
As mentioned in Section 4, it was derived by Stark \& Piran (1985, 1986)
and it refers to a simplified model of star, namely a naked 
compact core with a polytropic equation of state. 
At present, detailed information on the gravitational radiation emitted
in collapses of stars with more realistic
equations of state, and derived by a fully relativistic approach as in the 
Stark-Piran model, are not available. 

As far as the ``naked core" assumption is concerned, it should be 
mentioned that detailed studies on the structure of
stars that evolve into Type II
supernovae, show that quite a general pre-collapse
configuration is composed by a degenerate iron  core
surrounded by layers of matter, remnants of the earlier
nuclear burning processes  (Woosley \& Weaver 1995)
However, these layers  are not expected
to contribute significantly to the emitted radiation, because
they do not participate in the core bounce (see for example the
discussion in Section VII of Seidel 1991). 

The preminent features of the Stark-Piran energy spectrum are (see Figure 3):
\begin{itemize}
\item The presence of a peak at the frequency of the quasi-normal 
modes of the formed black hole.  
\item  The dependence of the efficiency on the fourth power of the angular 
momentum of the collapsing star.
\item  The presence of a further pronounced peak at a lower frequency, 
if the angular momentum is sufficiently high.
\end{itemize}
It is interesting to ask to what extent are these features general,
and how do they reflect in the structure of the quantities we 
compute, for instance  the strain amplitude, which is of direct
interest for gravitational detectors.
Gravitational signals emitted during
the collapse of naked cores to a black hole have been studied 
in several papers with a perturbative approach:  a spherically 
symmetric collapsing star is perturbed axisymmetrically, and the 
emitted radiation is computed by solving the linearized equations
(see Ferrari \& Palomba 1998, for a recent review).
The amount of  gravitational energy  
estimated by the perturbative approach
is obviously lower than  that predicted by Stark and Piran. 
However, the presence of the peak of the quasi-normal modes can be 
considered as a general feature, unless the dynamics of the 
fluid is dominated by very strong internal pressure or by very long bounces, 
which basically slow down the collapse to such an extent that the formed 
black hole has no residual mechanical energy to radiate away in 
its quasi-normal modes (Seidel 1991).  
The ``two-peak" structure of the energy spectrum can be mantained also 
if the star is non-rotating (Seidel 1991), if bounces occur during the 
collapse. The dependence of the emitted energy on the fourth power 
of the angular momentum is found also in the perturbative approach
(Cunningham, Price \& Moncrief 1980). In all cases the frequency scales
as the inverse of the black hole mass, and the energy radiated 
as the second power of the mass.
Thus, the features of the energy spectrum  we use to model each single
event are likely to reasonably represent a generic situation.

The strain amplitude we computed from the Stark-Piran energy spectrum
is shown in Figure 5. It exhibits a peak at some frequency
which is the reminiscent of the peak of the quasi-normal modes 
of the formed black holes. 
Since we do not know  the angular momentum distribution of the 
black holes, each  curve in that figure is obtained by assuming that 
all black holes that form have the same value of the rotation parameter, and 
the location of the peak of \op\sqrt{S_h} \cl
depends on that value. 
We can summarize the above discussion by saying that the strain 
amplitude associated to the background under consideration should 
present a peak at a frequency between $\sim$ 230 and 340 Hz,
whereas the expected maximum amplitude should be of order 
\op 1-8 ~\times 10^{-28}.\cl 
However, since our results are based on the assumption that each 
collapse takes place axisymmetrically the maximum strain amplitude may
be underestimated: non-axisymmetric collapse would be more 
efficient in producing gravitational waves.
The strain amplitude plotted in Figure 5 was 
derived for  the following set of parameters: 
$\alpha=0.1$, $M_p=25\msun$ and $M_u=125\msun$. 
The dependence on different choices for the
progenitor mass range ($M_p$, $M_u$)  was considered in section
6. The results are shown in Figure 8 and in Table 2.
For a lower value of the upper mass cutoff, the frequency of the maximum
is shifted by a factor $\sim 2.4$ towards higher frequencies whereas the
maximum amplitude is lowered by a factor $\sim 2.2$. Similarly, Figure 9
shows the effects of variating the fraction of the progenitor 
mass which participates to the collapse ($\alpha$). We can consider 
these curves as an indication of the uncertainties introduced by the
lack of consistent results in the literature. 
However, reasonable values of the parameters $\alpha$, 
$M_p$ and $M_u$ 
should not be too far from the ranges we have adopted. Thus, we may 
conclude that, according to
our study, the spectral strain amplitude produced by the background of 
supernova explosions leading to black hole formation should have a maximum at 
a few hundred Hz with an amplitude ranging between 
\op 10^{-28}\cl and \op 10^{-27}\cl Hz$^{-1/2}$. 
The corresponding closure density, \op \Omega_{GW},\cl has a
maximum amplitude ranging between \op 10^{-11}\cl and \op 10^{-10}\cl 
in the interval \op \sim 1.5-2.5\cl kHz. 
This is beyond the sensitivity that can be obtained by 
cross-correlating the
outputs of two gravity-wave detectors of the first generation experiments 
(VIRGO-VIRGO, LIGO-LIGO). However, for a constant frequency spectrum the
advanced LIGO detector pair aims at $\Omega_{GW} \sim 2 \times 10^{-10}$
(Allen \& Romano 1998), a level comparable to our signal. 

Finally, we have made a preliminary statistical analysis of our
background, e.g. in terms of the probability that the signal is above a given 
threshold. More details will be considered in a forthcoming paper. 
Let us stress, once again, that the smallness of the duty cycle,
which makes our stochastic background a non-continuous one, implies a 
non-trivial statistics of the signal arrival times which will be a 
distinctive feature of the process that can be in principle exploited 
to design optimal detection strategies. 

\section*{Acknowledgments} 

We would like to thank Piero Madau for kindly supplying 
the star formation rate evolution model and for stimulating 
discussions. 
Pia Astone, Enrico Cappellaro, Sergio Frasca, Cedric Lacey, Francesco Lucchin, 
Cristiano Palomba, Lucia Pozzetti and Roberto Turolla are also acknowledged 
for useful conversations and fruitful insights in various aspects of the
work. 
We thank the Italian MURST for partial financial support.

\label{lastpage}

\end{document}